\documentclass[aps,prx,twocolumn,superscriptaddress]{revtex4-1}

\usepackage{amsmath}
\usepackage{amssymb}
\usepackage{amsthm}
\usepackage{amsbsy}
\usepackage{mathrsfs}
\usepackage{graphicx}
\usepackage{enumerate}
\usepackage{xcolor}   
\usepackage{array}
\usepackage{enumitem}
\usepackage{multirow}
\usepackage{float}
\usepackage[utf8]{inputenc}
\usepackage[T1]{fontenc}
\usepackage{enumerate}
\usepackage[colorlinks=false]{hyperref}
\usepackage{soul} 
\usepackage{changes} 
\usepackage{setspace} 

\usepackage{lipsum}

\makeatletter
\renewcommand*{\@biblabel}[1]{\hfill#1.}
\makeatother

\usepackage{pgfplots}
\pgfplotsset{compat=newest}
\usetikzlibrary{plotmarks}
\usetikzlibrary{arrows.meta}
\usepgfplotslibrary{patchplots}
\usepackage{grffile}

\theoremstyle{plain}
\newtheorem{thm}{Theorem}
\newtheorem{ass}{Assumption}

\pgfplotsset{compat=newest} 
\pgfplotsset{plot coordinates/math parser=false}

\newcommand{\ket}[1]{\ensuremath{|{#1}\rangle}}
\newcommand{\bra}[1]{\ensuremath{\langle{#1}|}}

\newcommand{\proj}[2]{\ensuremath{|{#1}\rangle\langle{#2}|}}

\newcommand{\vecP}{\vec{\wp}}
\newcommand{\Pext}{\wp_\text{\tiny Ext}}
\newcommand{\ie}{i.e.}
\newcommand{\eg}{e.g.}

\renewcommand{\figurename}[1]{Fig.} 

\definecolor{red}{rgb}{1,0,0}
\definecolor{blue}{rgb}{0,0,0.8}
\definecolor{maroon}{rgb}{0.65,0,0}
\definecolor{ngreen}{rgb}{0.3,0.6,0.2}
\definecolor{pink}{rgb}{1.0,0.3,0.7}
\definecolor{golden}{rgb}{0.8,0.6,0.1}
\definecolor{teal}{rgb}{0.1,0.55,0.55}

\usepackage{mfirstuc}
 
\makeatletter
\renewcommand\@make@capt@title[2]{%
  \@ifx@empty\float@link{\@firstofone}{\expandafter\href\expandafter{\float@link}}%
   {\textbf{#1}}\@caption@fignum@sep#2\quad
}%

\makeatother

\begin{document}

\title{A strong no-go theorem on the Wigner’s friend paradox}

\author{Kok-Wei Bong}
\email{these authors contributed equally to this work}
\affiliation{Centre for Quantum Computation and Communication Technology (Australian Research Council), Centre for Quantum Dynamics, Griffith University, Brisbane, QLD 4111, Australia}

\author{An\'ibal Utreras-Alarc\'on}
\email{these authors contributed equally to this work}
\affiliation{Centre for Quantum Computation and Communication Technology (Australian Research Council), Centre for Quantum Dynamics, Griffith University, Brisbane, QLD 4111, Australia}

\author{Farzad Ghafari}
\affiliation{Centre for Quantum Computation and Communication Technology (Australian Research Council), Centre for Quantum Dynamics, Griffith University, Brisbane, QLD 4111, Australia}

\author{Yeong-Cherng~Liang}
\affiliation{Department of Physics and Center for Quantum Frontiers of Research \& Technology (QFort), National Cheng Kung University, Tainan 701, Taiwan}

\author{Nora Tischler}
\email{n.tischler@griffith.edu.au}
\affiliation{Centre for Quantum Computation and Communication Technology (Australian Research Council), Centre for Quantum Dynamics, Griffith University, Brisbane, QLD 4111, Australia}

\author{Eric G. Cavalcanti}
\email{e.cavalcanti@griffith.edu.au}
\affiliation{Centre for Quantum Computation and Communication Technology (Australian Research Council), Centre for Quantum Dynamics, Griffith University, Gold Coast, QLD 4222, Australia}

\author{Geoff J. Pryde}
\affiliation{Centre for Quantum Computation and Communication Technology (Australian Research Council), Centre for Quantum Dynamics, Griffith University, Brisbane, QLD 4111, Australia}

\author{Howard M. Wiseman}
\affiliation{Centre for Quantum Computation and Communication Technology (Australian Research Council), Centre for Quantum Dynamics, Griffith University, Brisbane, QLD 4111, Australia}

\begin{abstract}
{Does quantum theory apply at all scales, including that of observers? New light on this fundamental question has recently been shed through a resurgence of interest in the long-standing Wigner’s friend paradox. This is a thought experiment addressing the quantum measurement problem---the difficulty of reconciling the (unitary, deterministic) evolution of isolated systems and the (non-unitary, probabilistic) state update after a measurement. Here, by building on a scenario with two separated but entangled friends introduced by Brukner, we prove that if quantum evolution is controllable on the scale of an observer, then one of `No-Superdeterminism', `Locality' or `Absoluteness of Observed Events'---that every observed event exists absolutely, not relatively---must be false. We show that although the violation of Bell-type inequalities in such scenarios is not in general sufficient to demonstrate the contradiction between those three assumptions, new inequalities can be derived in a theory-independent manner, that are violated by quantum correlations. This is demonstrated in a proof-of-principle experiment where a photon’s path is deemed an observer. We discuss how this new theorem places strictly stronger constraints on physical reality than Bell’s theorem.}
\end{abstract}
\maketitle

\section{}
Wigner's friend \cite{Wigner} is a thought experiment that illustrates what is perhaps the thorniest foundational problem in quantum theory: the measurement problem~\cite{Schlosshauer2004, Leggett2005}. In the thought experiment, we consider an observer (the ``friend'') who performs a measurement on a quantum system.
In accordance with the state update rule, the friend assigns the eigenstate corresponding to their observed outcome to the measured system. The friend is assumed to be inside an isolated laboratory that can be coherently controlled by a second experimenter, Wigner, who is capable of performing arbitrary quantum operations on the friend's laboratory and all of its contents. Although this may be possible, in principle, it would be a truly Herculean task if the friend were a macroscopic observer like a human, as we have chosen for our illustrations and discussions below. For this reason Wigner is often called a \emph{superobserver}. However, there is good reason to think that quantum mechanics would allow control of the type required if the friend were an artificial intelligence algorithm in a simulated environment running in a large quantum computer. Wigner describes the laboratory and all of its contents as a unitarily evolving quantum state, in accordance with the rule for state evolution applicable to isolated systems. The case when the friend’s system is prepared in a superposition state leads to an apparent contradiction between the friend’s perspective and that of Wigner, who does not ascribe a well-defined value to the outcome associated with his friend’s observation. For a more in-depth description of the Wigner's friend thought experiment, see the Supplementary Information, Sec.\ A.

Although decoherence can ``save the appearances'' by explaining the suppression of quantum effects at the macroscopic level, it cannot solve the measurement problem: ``We are still left with a multitude of (albeit individually well-localized quasiclassical) components of the wave function, and we need to supplement or otherwise to interpret this situation in order to explain why and how single outcomes are perceived''~\cite{Schlosshauer2004}. Proposed resolutions have radical implications: they either reject the idea that measurement outcomes have single, observer-independent values~\cite{Everett1957, Rovelli1996, Fuchs2013, Mermin2014}, or postulate faster-than-light~\cite{Bohm1, Bohm2} or retrocausal effects~\cite{Price2008, tHooft2007} at a hidden-variable level. Alternatively, some theories postulate mechanisms to avoid macroscopic superpositions, such as modifications to unitary quantum dynamics~\cite{Bassi2003} or gravity-induced collapse~\cite{Penrose1996}. Here we rigorously demonstrate that radical revisions of such types are in fact required.

Our work is inspired by the recent surge of renewed interest in the Wigner's friend problem \cite{BruknerLF,BruknerBook, FR, Proietti2019,Formalisms,Healey2018,BruknerComment}. In particular, Brukner~\cite{BruknerLF} introduced an extended Wigner's friend scenario (EWFS) with two spatially separated laboratories, each containing a friend, accompanied by a superobserver who can perform various measurements on their friend's laboratory. Each friend measures half of an entangled pair of systems, establishing correlations between the results of the superobservers' subsequent measurements. 

In the context of this EWFS, Brukner~\cite{BruknerLF,BruknerBook,BruknerComment} considered three assumptions, namely: \textit{{freedom of choice}}, \textit{{locality}} (in the sense of ``parameter independence''~\cite{Shimony1984}) and \textit{{observer-independent facts}} (OIF). The last of these means that propositions about all observables that might be measured (by an observer or a superobserver) are ``assigned a truth value independently of which measurement Wigner performs’’~\cite{BruknerLF}. 

In other words, the OIF assumption is equivalent to the assumption of {\em Kochen-Specker noncontextuality}~\cite{KS1967, Liang2011} ({KSNC}). From these assumptions, Brukner derived a Bell inequality for the correlations of the superobservers' results, which could be violated in quantum mechanics (if the superobservers could suitably manipulate the quantum state of the observers). A recent six-photon experiment~\cite{Proietti2019}, using a set-up where the role of each friend is played by a single photon, successfully violated such a Bell inequality derived from Brukner's assumptions.

Although the EWFS background for this result was novel, the derived Bell inequality can be obtained from the assumptions of {freedom of choice} and {KSNC}, without considering the friends' observations, and without using ``{locality}'' (which follows from Bell's stronger notion of {local causality}~\cite{Causarum}, which in turn follows from {KSNC} in any Bell scenario~\cite{Cavalcanti2018}). Furthermore, the Kochen-Specker theorem~\cite{KS1967} already establishes that KSNC + ``{freedom of choice}'' leads to contradictions with quantum theory. As discussed in refs.~\cite{Healey2018,BruknerComment,HealeyReply}, this casts doubt on the implications of Brukner's theorem with regard to any assumption specifically about the objectivity of the friends’ observations---one can respond to Brukner's theorem simply by maintaining that ``\emph{unperformed} experiments have no results''~\cite{Peres1978}. 

Nevertheless, there is a subtle but important difference between a standard Bell scenario in which one of two incompatible observables are chosen at random to be measured by each party and the scenario introduced by Brukner. In the latter, in one out of four experimental runs all four observables involved in the experiment are being measured---one by each observer in the scenario. This suggests that the counterfactual reasoning in the {OIF/KSNC} assumption could be avoided by replacing it with a suitable weaker assumption. Indeed, Brukner discusses a weaker 
assumption---``that Wigner’s and Wigner’s friend's facts coexist''---before settling on  ``The assumption of `observer-independent facts' [which] is a stronger condition''~\cite{BruknerLF}.

In this Article we derive a new theorem, based on the intuition in the preceding paragraph around Brukner's EWFS. It uses metaphysical assumptions (i.e.~assumptions \emph{about} physical theories) that are strictly weaker than those of Bell's theorem or Kochen-Specker contextuality theorems, and thus opens a new direction in experimental metaphysics. Our first two assumptions are, as per Brukner, ``{freedom of choice}'' (which we make more formal using the concept of ``{No-Superdeterminism}'' defined in Ref.\cite{Causarum}), and {Locality} (in the same sense as Brukner; see also Ref.\cite{Causarum}). Our third assumption is {\em {Absoluteness of Observed Events}} (AOE), which is that an observed event is a real single event, and not relative to anything or anyone. Note that capitalization is used for assumptions formally defined in this paper.

\begin{figure}[h]
    \centering
    \includegraphics[width=0.7\linewidth]{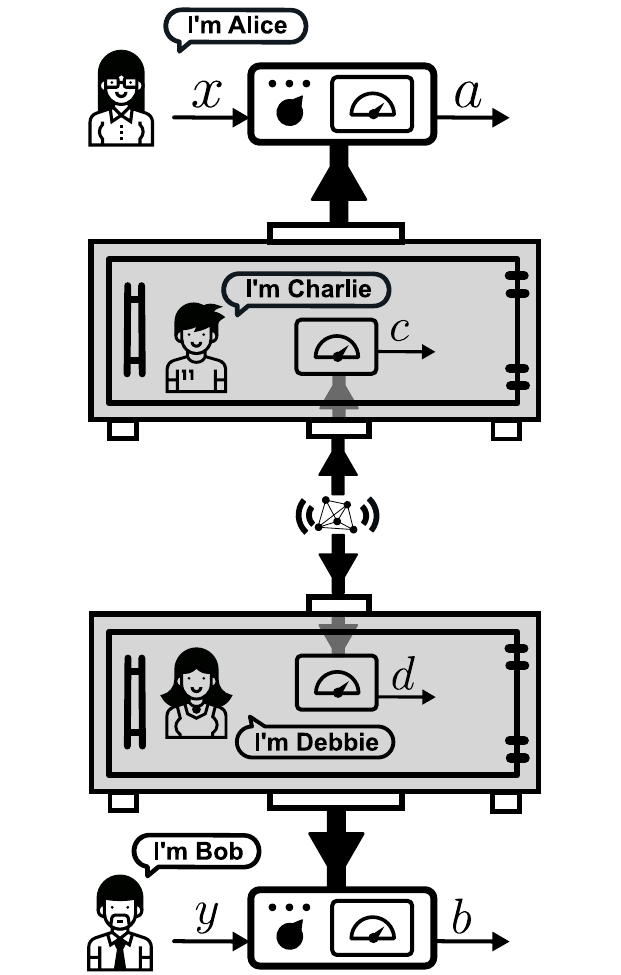}
    \caption{Concept of the extended Wigner's friend scenario. {The friends,} Charlie and Debbie, measure a pair of particles prepared in an entangled state, producing outcomes labelled $c$ and $d$, respectively (from their perspective). {The superobservers,} Alice and Bob, perform space-like separated measurements labelled $x$ and $y$, with outcomes labelled $a$ and $b$, on the entire contents of the laboratories containing Charlie and Debbie, respectively.}
    \label{fig:concept_fig}
\end{figure}

Unlike OIF, AOE makes no claim about hypothetical measurements that were not actually performed in a given run. Furthermore, AOE is necessarily (though often implicitly) assumed even in standard Bell experiments~\cite{Causarum}. For convenience, we will call the conjunction of these three 
assumptions \textit{{Local Friendliness}} (LF). This enables us to state our theorem: 
\begin{thm}\label{theorem_1}
    If a superobserver can perform arbitrary quantum operations on an observer and its environment, then no physical theory can satisfy Local Friendliness.
\end{thm}
By a ``physical theory'' we mean any theory that correctly predicts the correlations between the outcomes observed by the superobservers Alice and Bob (see Fig.~\ref{fig:concept_fig}), who can communicate after their experiments are performed and evaluate those correlations. The proof of Theorem 1 proceeds by showing that LF implies a set of constraints on those correlations (that we call \emph{LF inequalities}) that can, in principle, be violated by quantum predictions for an EWFS scenario. Thus, like Bell's theorem and Brukner's theorem, our theorem is theory-independent---we use (like Bell and Brukner) quantum mechanics as a guide for what may be seen in experiments, but the metaphysical conclusions hold for any theory if those predictions are realized in the laboratory. (This is unlike the theorem of Ref.~\cite{FR}, which is a statement about the standard theory of quantum mechanics.) Note also that, unlike in Brukner's theorem, all three assumptions going into LF are essential for the theorem, and so are the friends' observations.

For the specific EWFS Brukner considered---involving two binary-outcome measurement choices per superobserver---the  set of correlations allowed by our LF assumption is identical to the set allowed by the assumptions of Bell's theorem, commonly referred to as the local hidden variable (LHV) correlations. However, in general, LF and LHV do {\em not} give identical constraints.  
Indeed, already for a slightly more complicated EWFS with three binary-outcome measurement choices per superobserver, we show that the set of LF correlations is a strict superset of the set of LHV correlations. Moreover, it is possible for quantum correlations to violate a Bell inequality (an inequality bounding the set of LHV correlations) while satisfying all of the LF inequalities. We also prove that the new LF inequalities we derive can nevertheless be violated by quantum correlations. We demonstrate these facts in an experimental simulation where the friends are represented by photon paths. 

We now proceed to explain the EWFS in more detail, before presenting our results and discussing their implications.

\subsection{The extended Wigner's friend scenario}\label{sec:extendedWigner}

\begin{figure}[h]
    \centering
    \includegraphics[width=0.8\linewidth]{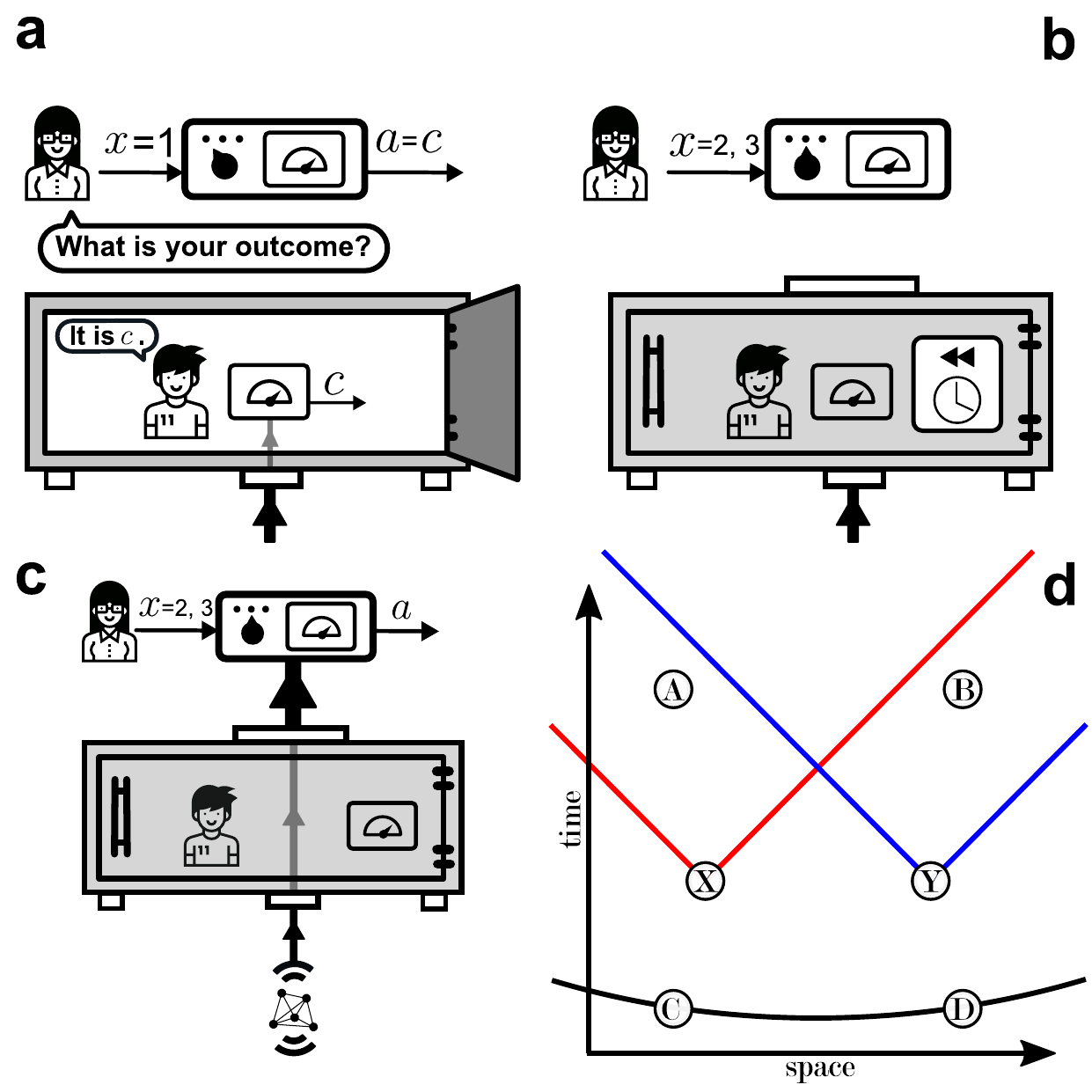}
    \caption{{A specific} bipartite Wigner's friend experiment. \textbf{a,} When $x=1$, Alice opens Charlie's laboratory and asks him his outcome. \textbf{b,} Alternatively, for $x = 2,3$, she may restore the laboratory to a previous state. \textbf{c,} She then proceeds to ignore Charlie, and performs a measurement directly on the particle. \textbf{d,} Space-time diagram illustrating the time ordering of the events within the experiment---C (D) is Charlie's (Debbie's) measurement, X (Y) is the event of Alice's (Bob's) choice of measurement setting, A (B) is Alice's (Bob's) measurement.}
    \label{fig:concept_comic}
\end{figure}

Let us consider the bipartite version of the Wigner's friend experiment that was introduced by Brukner, involving two superobservers, Alice and Bob, and their respective friends, Charlie and Debbie (Fig.\ \ref{fig:concept_fig}). Charlie and Debbie each have one particle from an entangled pair, and make a measurement on it, yielding outcomes $c$ and $d$, respectively.

In each iteration of the experiment, Alice and Bob randomly and independently choose one out of $N\ge2$ measurements to be performed in space-like separated regions subsequent to a space-like hypersurface containing the measurements of both Charlie and Debbie, as shown in Fig.~\ref{fig:concept_comic}d. The settings are respectively labelled $x \in \{1,...,N\}$ and $y \in \{1,...,N\}$, with corresponding outcomes $a$ and $b$ (we do not assume anything about the number of possible outcomes at this stage). For the specific EWFS depicted in Fig.~\ref{fig:concept_comic}, if $x=1$, Alice simply opens Charlie's laboratory and directly asks him for his outcome $c$, then assigns her own outcome as $a = c$, as shown in Fig.~\ref{fig:concept_comic}a.

For $x\in\{2,...,N\}$, Alice performs a different measurement on Charlie's laboratory as a whole. In particular, we will consider measurements such that Alice restores the laboratory to a previous state, thereby erasing Charlie's memory (Fig.~\ref{fig:concept_comic}b), and then proceeds to measure the particle directly (Fig.~\ref{fig:concept_comic}c). Bob and Debbie operate in a similar fashion.

From this experiment we can measure (as frequencies) the empirical probabilities $\wp(ab|xy)$, using only the information available at the end of the experiment, namely, the values for $a$, $b$, $x$ and $y$. Unless $x=1$, all records for the value of $c$ are erased when Alice performs her measurement, so in general that information cannot be accessed at the end of the experiment, and likewise with the value of $d$ on Bob's side. A detailed quantum mechanical description of the EWFS is provided in Methods.

\subsection{Formalization of the {Local Friendliness} assumptions}\label{sec:formalization}

Within a bipartite Wigner's experiment, what constraints do the LF assumptions imply for the probabilities $\wp(ab|xy)$ observed by Alice and Bob for outcomes $a$ and $b$, given settings $x$ and $y$? To determine this rigorously we need to formalize our three assumptions. 

\begin{ass}[\textbf{{Absoluteness of Observed Events} (AOE)}] \label{defin:AF}
An observed event is a real single event, and not relative to anything or anyone.
\end{ass}

In an EWFS, the assumption of AOE implies that \emph{in each run of the experiment}---\ie~given that Alice has performed measurement $x$ and Bob has performed measurement $y$ on some pair of systems---there exists a well-defined value for the outcome observed by each observer, that is, for $a$, $b$, $c$ and $d$. Formally, this implies that there exists a theoretical joint probability distribution $P(abcd|xy)$ from which the empirical probability $\wp(ab|xy)$ can be obtained while also ensuring that the observed outcomes for $x,y=1$ are consistent between the superobservers and the friends.
\begin{itemize}
\item \textit{{AOE} (in the EWFS of Fig.~\ref{fig:concept_comic}):} 

$\exists~P(abcd|xy) \; \mathrm{s.t.}$
\begin{enumerate}
    \item[i)] $\wp(ab|xy)=\Sigma_{c,d}P(abcd|xy)$ $\forall \, a,b,x,y$,
    \item[ii)]  $P(a|cd,x=1,y)=\delta_{a,c}$ $\forall \, a,c,d,y$,
    \item[iii)] $P(b|cd,x,y=1)=\delta_{b,d}$ $\forall \, b,c,d,x$.
\end{enumerate}
\end{itemize}

Here we \emph{do not} assume that all statements about results have truth values independently of which measurement `Wigner' (whom we call Alice) performs. Instead, 
the assumption of {AOE} only entails assigning truth values to propositions about \emph{observed} outcomes. 
In particular, Alice's measurement outcome $A_x$ (which in our notation corresponds to the value of $a$ when she performs measurement labelled by $x$) for $x\neq 1$ has a value \emph{only when she performs that measurement}.
However, $A_1$ is different in that it has a value even when $x\ne1$, because it is encoded in $c$, which is \emph{actually} measured by Charlie in every run. All this is in keeping with Peres' dictum ``unperformed experiments have no results''~\cite{Peres1978}; AOE is the assumption that \emph{performed} experiments have observer-independent (that is, absolute) results. 

The {No-Superdeterminism} assumption is a formalization of the assumption of ``freedom of choice'' used in derivations of Bell inequalities. It is the assumption that the experimental settings can be chosen freely, that is, uncorrelated with any relevant variables prior to that choice. For added clarity, here we formulate it, following ref.~\cite{Causarum} as follows.

\begin{ass}[\textbf{{No-Superdeterminism} (NSD)}] Any set of events on a space-like hypersurface is uncorrelated with any set of freely chosen actions subsequent to that space-like hypersurface.
\end{ass}

In the EWFS, this implies that $c$ and $d$ are independent of the choices $x$ and $y$:
\begin{itemize}
\item \textit{{NSD} (in the EWFS and under Assumption 1):} 

$P(cd|xy)=P(cd)$ $\forall \, c,d,x,y$.
\end{itemize}

Finally, the assumption of {Locality} prohibits the influence of a local setting (such as $x$) on a distant outcome (such as $b$). It is the assumption that Bell, in 1964~\cite{Bell1964}, and many others subsequently, also called ``{locality}''~\cite{Causarum}, and which Shimony called ``parameter independence'' \cite{Shimony1984}; that is, in the formalization of ref.~\cite{Causarum}, the following assumption. 
\begin{ass}[\textbf{{Locality} (L)}] The probability of an observable event $e$ is unchanged by conditioning on a space-like-separated free choice $z$, even if it is already conditioned on other events not in the future light-cone of $z$.
\end{ass}
In the EWFS, this
implies:

\begin{itemize}
\item \textit{{L} (in the EWFS and under Assumption 1):}

$P(a|cdxy)=P(a|cdx)$ $\forall \, a,c,d,x,y$,\\ 
$P(b|cdxy)=P(b|cdy)$ $\forall \, b,c,d,x,y$.
\end{itemize}

Note that one could alternatively formulate Assumptions 2 and 3 as a single, equivalent assumption, which has previously been coined ``local agency'' in the context of Bell's theorem \cite{Causarum}. 
Within the definitions of L and  NSD, $c,d$ play the formal role of the hidden variables $\lambda$ in the usual derivation of Bell inequalities. However, we emphasize again that those correspond to \emph{observed} events, and note that we make no assumption about hidden variables predetermining all measurement outcomes.

We call the set of correlations $\wp(ab|xy)$ that satisfy Assumptions 1--3 the LF correlations.

\subsection{Properties of LF correlations}

\begin{table*}[t]
\onecolumngrid
\centering
\begin{tabular}{|c|c|c|c|}\hline
\bf Label       & \; \bf Measurement Settings & \bf \;LF inequality?\;  &\;\bf Bell facet?\;\\\hline
Brukner     &($1\,i,1\,j$)                                      & Yes                &Yes\\
Semi-Brukner&($1\,i,2\,3$)                                      & Yes                &Yes\\
Bell non-LF &($2\,3,2\,3$)                                      & No                 &Yes\\
$I_{3322}$  &($1\,2\,3,1\,2\,3$)                                & Yes                &Yes\\
Genuine LF &($1\,2\,3,1\,2\,3$)                                & Yes                &No\\
\hline
\end{tabular}
\caption{\small Categories of inequalities for three binary-outcome measurement settings per party. The column Measurement Settings refers to the settings that appear in each inequality, with $i,j\in\{2,3\}$. The third column specifies whether it is an LF inequality, and the fourth column specifies whether it is a facet of the Bell polytope. Each category represents inequalities with the same form up to arbitrary relabeling of measurement settings (for $i,j \neq 1$), outcomes, and parties. The labels referring to each inequality are: ``Genuine LF'' for inequalities that are not facets of the LHV polytope for this scenario, ``$I_{3322}$'' for a type of Bell facets for the case of three binary-outcome measurement settings per party~\cite{CollinsGisin}, and ``Brukner'', ``Semi-Brukner'' and ``Bell non-LF'' are inequivalent classes of CHSH-type inequalities~\cite{CHSH}. ``Brukner'' inequalities are the type of inequalities considered by Brukner~\cite{BruknerLF}. A ``Semi-Brukner'' inequality has a simpler experimental realization than a ``Brukner'' inequality, as it only requires one of the parties to measure a friend (setting 1). ``Bell non-LF'' inequalities are Bell facets, but unlike the other categories, are not facets of LF.}
\label{table}
\twocolumngrid
\end{table*}

\begin{figure}
    \centering
    \includegraphics[width=0.7\linewidth]{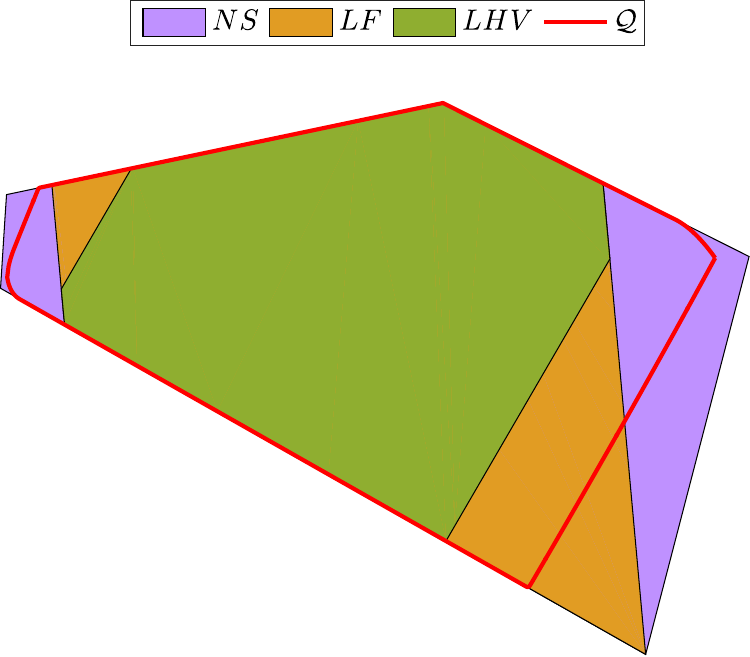}
\caption{A two-dimensional slice of the space of correlations, illustrating the correlations discussed in this work. The solid areas depict a hierarchy of models:  LHV {~\cite{Bell1964}} correlations (green) are a subset of LF correlations (green and orange), which in turn are a subset of no-signalling{~\cite{Barrett05}} correlations (NS, green and orange and purple). 
The red line bounds the correlations allowed by quantum theory on this slice. Note that, although the set of quantum correlations include the LHV set, it does not include the LF set. Further details of this plot are discussed in the Supplementary Information  (Sec.\ C).}
    \label{fig:2dslice}
\end{figure}

Our key findings about the properties of LF correlations are as follows. (a) LF correlations are a superset of LHV correlations, and in general a strict superset, as we will show quantitatively in the next section. (b) LF correlations can always be characterized by a finite set of inequalities. (c) For $N=2$ measurement settings and any number of measurement outcomes, LF correlations are the same as LHV correlations. (d) For $N=3$ measurement settings and $O=2$ outcomes, we fully characterize the LF correlations by deriving the associated inequalities and we show that they are a strict superset of LHV correlations (as illustrated in Fig.\ \ref{fig:2dslice}). We provide the derivations to these results in the Methods.

For $N=3$ measurement settings and $O=2$ outcomes, the set of LF correlations is a polytope with 932 facets. The facets can be grouped into nine inequivalent classes, each represented by a different inequality (provided in the Methods). These classes can be further grouped into categories, according to the measurement settings involved, and whether the facets are Bell facets~\cite{BrunnerRMP}. 
In Table \ref{table}, we list the categories of LF facets, ignoring all positivity facets, that is, the constraints that probabilities cannot be negative.

\subsection{Quantum violations}\label{sec:Quantum violations}

\begin{figure}
\centering
\resizebox{0.75\linewidth}{!}{\includegraphics{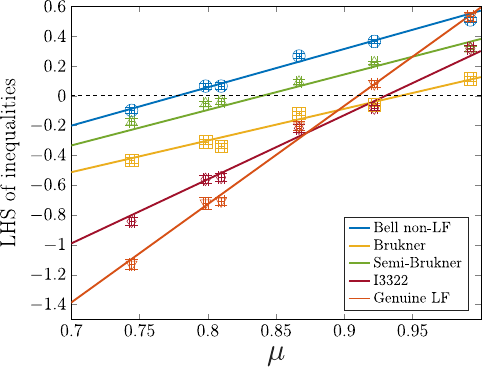}}
\caption{\small Results for the left-hand sides of Bell and LF inequalities for different quantum states. The parameter $\mu$ is the pure-state fraction of the quantum state in Eq.~(\ref{quasi-W}). The measurements and inequalities considered are provided in the Methods, using the labels introduced in Table \ref{table}. The dashed line in the plot represents the bound above which a violation occurs. The solid lines are theory predictions and the symbols represent experimental data. The uncertainties for the data points represent $\pm 1$ standard deviations, calculated from Monte Carlo simulations using 100 samples of Poisson-distributed photon counts. Reproduced from ref.\ \cite{Proceedings}, Society of Photo‑Optical Instrumentation Engineers (SPIE).}
\label{fig:results}
\end{figure}

We now search for quantum violations of the LF inequalities. To demonstrate that the set of LF correlations is strictly larger than the LHV correlations, we seek a state and measurement choices such that a violation of a ``Bell non-LF'' inequality is exhibited without a violation in any of the LF inequalities. For 
experimental convenience, we consider two-qubit photon polarization states of the form
\begin{equation}
    \rho_\mu = \mu \proj{\Phi^-}{\Phi^-} +\dfrac{1 - \mu}{2} ( | HV \rangle \langle HV | + | VH \rangle \langle VH | ) \,
    \label{quasi-W}
\end{equation}
where $ \ket{\Phi^-} = (\ket{HV} - \ket{VH})/\sqrt{2}$,  $0\leq \mu \leq 1$, and $H$ and $V$ denote horizontal and vertical polarizations, respectively.

In Fig.\ \ref{fig:results} we display quantum violations for inequalities of all the categories in Table \ref{table} for states $\rho_\mu$. The specific inequalities and measurements considered are described in the Methods. Each of the inequalities considered is violated by some $\rho_\mu$. In addition, we determine the strongest violations of the Genuine LF inequalities allowed in quantum theory; those results are provided in the Supplementary Information (Sec.\ B).

In summary, if quantum measurements can be coherently performed at the level of observers, quantum mechanics predicts the violation of the LF inequalities in EWFSs. This proves Theorem 1.

\subsection{Experiment}

\begin{figure*}[t]
\onecolumngrid
\centering
  \includegraphics[width=0.75\linewidth]{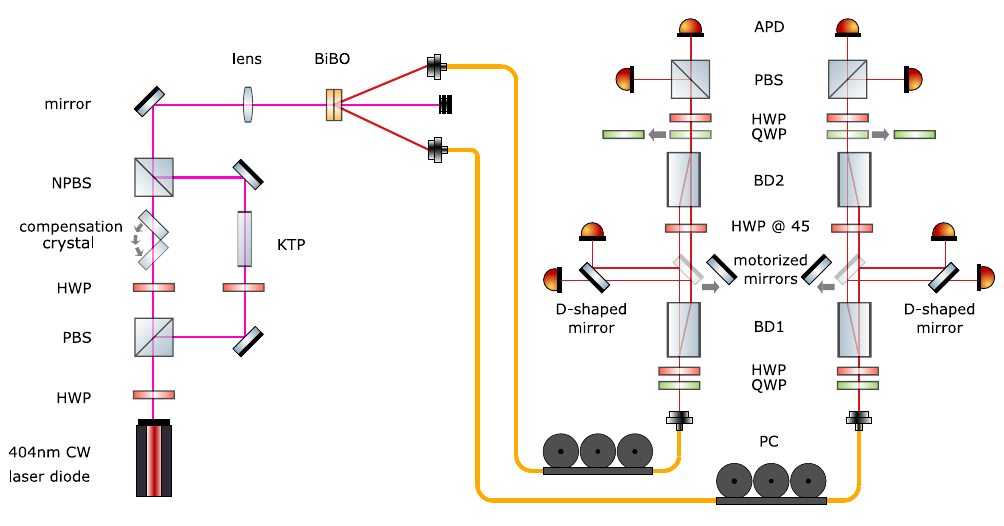}  
\caption{\small Experimental setup. The source is depicted on the left-hand side, and the measurement section on the right-hand side. The desired quantum state is generated via type-I spontaneous parametric down-conversion using two orthogonally oriented BiBO crystals. The pump beam for the down-conversion process is a mixture of a decohered state that is obtained from the long arm of the interferometer and a diagonally polarized state from the short arm. The measurement section allows for tomography to be carried out when the motorized mirrors are removed and the photons traverse the beam displacer (BD) interferometers. Alice and Bob perform projective measurements when the quarter-waveplates (QWP) of the tomography stages are removed. Alternatively, they can ask Charlie and Debbie for their respective measurement outcomes by sliding in the motorized mirrors, using the fact that the projective measurements of their friends correspond to the beam paths inside the interferometers. Abbreviations used: Non-polarizing beam splitter (NPBS), potassium titanyl phosphate (KTP), half-wave plate (HWP), avalanche photodiode (APD), polarization control (PC), polarizing beam splitter (PBS). Reproduced from ref.\ \cite{Proceedings}, Society of Photo‑Optical Instrumentation Engineers (SPIE).}
\label{fig:expt fig}
\twocolumngrid
\end{figure*}

We study the EWFS with three measurement settings ($N=3$) in an experiment where the systems distributed between the two laboratories are polarization-encoded photons, the friends are photon paths within the set-up and the measurements by the superobservers are photon-detection measurements. Because the qubit composed of the two photon paths that represents each of our friends would not typically be considered a macroscopic, sentient observer as originally envisioned by Wigner, our experiment is best described as a proof-of-principle version of the EWFS. The experiment lets us demonstrate the key properties of LF inequalities and its results generalize provided that quantum evolution is, in principle, controllable on the scale of an observer. A fully rigorous demonstration that the LF assumptions are untenable would require, in addition to a more plausible ``observer'', implementing shot-by-shot randomized measurement settings and closing separation, efficiency, and freedom-of-choice loopholes, similarly to the case of Bell inequality violations \cite{Giustina2015, Hensen2015, Shalm2015}.

Our experimental set-up, which comprises a photon source and a measurement section, is illustrated in Fig.~\ref{fig:expt fig}. The photon source, shown in the left half of Fig.~\ref{fig:expt fig}, is designed to generate the quantum state $\rho_{\mu}$ of Eq.~(\ref{quasi-W}) with a tunable $\mu$ parameter. Details about this spontaneous parametric down-conversion source are provided in the Methods.

The measurement section of the experimental set-up, shown in the right half of Fig.~\ref{fig:expt fig}, consists of two copies of an apparatus, one belonging to Alice and Charlie, and the other to Bob and Debbie. The measurement section serves two purposes. The first is to perform quantum state tomography in order to characterize the generated quantum state, as detailed in the Methods and the Supplementary Information (Sec.\ D). The second purpose is to perform the measurements of the four observers, Charlie, Debbie, Alice and Bob. The friend's projective polarization measurement result is encoded in the photon path after the QWP, HWP, and BD1. Alice and Bob can perform different positive-operator valued measures (POVMs) on their respective system+friend, which depend on their measurement settings and are described in the Methods.

The experimental results are shown in Fig.\ \ref{fig:results}. The $\mu$ values cover the full range of interest, from none of the inequalities being violated (at low $\mu$), to the violation of all inequalities (at high $\mu$). The experimental data demonstrate the sequential violations of the Bell non-LF, semi-Brukner, and genuine LF inequalities. The data points corresponding to $\mu=0.80$ and $\mu=0.81$ are of particular significance, as they demonstrate that it is possible to violate Bell inequalities without violating any LF inequalities. (We can be confident of this because we verified that none of the 932 LF inequalities are violated.)

This means that the correlations consistent with LF assumptions are a superset of the correlations consistent with an LHV model. The case of $\mu=0.87$ is the first of the plotted data sets where a contradiction with the LF assumptions occurs, through the first violation of an inequality associated with LF. Finally, the two highest $\mu$ values verify that the genuine LF inequality can also be violated. All the experimental data points, except for the case of $\mu=0.81$, are at least two standard deviations away from 0, thus attesting the violation or non-violation of the inequalities with statistical significance. This covers all the 
regions we show in terms of \mbox{(non-)violation} of different inequalities, because the data set at $\mu=0.81$ belongs to the same region as $\mu=0.80$. Along with the experimental data, the results predicted for the design measurement directions and input states of Eq.\ (\ref{quasi-W}) are shown by the solid lines. However, because the inequalities are  device-independent, 
our conclusions are independent of which states and measurement directions were actually employed in the experiment.

\subsection{Implications of violating LF inequalities }

It is interesting to compare the assumptions that go into the LF no-go theorem with those for Bell's theorem.  First, we note that the AOE assumption is implicit in the derivation of Bell inequalities (see ref.~\cite{Causarum} for a derivation in which it is explicitly included as ``macroreality''). If, as is common, we also formulate Bell's theorem using the other two assumptions of LF, namely NSD and L,
then an additional assumption is 
required. The minimal extra assumption required is ``outcome independence'' \cite{Shimony1984}, which in the bipartite scenario is the requirement that $P(a|bxy\lambda)=P(a|xy\lambda)$, $P(b|axy\lambda)=P(b|xy\lambda)$ $\forall a,b,x,y,\lambda$ (c.f.~the definition of L in section Formalization of the {Local Friendliness} Assumptions). Hence, the LF assumption is strictly weaker than the set of  assumptions for Bell inequalities. Thus, the conclusions we could derive from an empirical violation of the LF inequalities are \emph{strictly stronger}. 

One popular way to accommodate the violation of Bell inequalities is to reject outcome independence (which is violated by operational quantum theory~\cite{Causarum}) while maintaining L and NSD. 
Our theorem shows that this strategy does not extend to the EWFS. If the LF inequalities were violated empirically, then, to maintain L and NSD, one would have to reject AOE.

It is important to 
keep in mind that it is much harder to satisfy the conditions for an experimental violation of the LF inequalities than of Bell inequalities. A fully convincing demonstration would require a strong justification for the attribution of a ``fact'' to the friend's measurement. This, of course, depends on what counts as an ``observer'' (and as a ``measurement''). Because conducting this kind of experiment with human beings is physically impractical, what do we learn from experiments with simpler ``friends''?

Wigner's own conclusion from his thought experiment was that the collapse of the wave function should happen at least before it reaches the level of an ``observer''. The concept of an ``observer'', however, is a fuzzy one. Objective collapse theories~\cite{Bassi2003} attempt to restore the absolute reality of observed events by postulating modifications to the quantum dynamics to guarantee that collapse occurs before a quantum superposition reaches the macroscopic level. In other words, this resolution requires observed events to correspond to sufficiently macroscopic irreversible physical processes. In that case, the LF inequalities would not be violated with actual observers. Clearly, our experiment (and that of ref.~\cite{Proietti2019}) did not probe collapse theories.  Therefore, an open possibility is that the LF assumptions are valid, but that nature forever forbids the observation of violation of LF inequalities with observers, whether because of objective collapse or some other limitation on coherent quantum control.

A challenge to the above resolution of the LF no-go theorem could come from experiments involving AI (artificial intelligence) agents in a quantum computer. If universal quantum computation and strong AI are both physically possible, it should be possible to realize quantum coherent simulations of an observer and its (virtual) environment, and realize an extended Wigner's friend experiment. The experiment can even be conducted with a single friend, which would already allow testing Semi-Brukner inequalities (\ref{eq:half-Brukner}).
Towards the goal of challenging the LF no-go theorem, experiments can test agents of increasing complexity; an experimental violation of LF inequalities with a given class of physical systems as ``friends'' implies that either the LF assumptions are false, or that class of friends is not an ``observer''.

Among interpretations of quantum mechanics that allow, in principle, the violation of LF inequalities, Theorem 1 can be accommodated in different ways. Interpretations that reject AOE include QBism~\cite{Fuchs2013, Mermin2014}, the relational interpretation~\cite{Rovelli1996}, and the many-worlds interpretation~\cite{Everett1957}. Bohmian mechanics~\cite{Bohm1, Bohm2} violates L but not the other assumptions. There are some advocates for
giving up NSD (either due to retrocausality~\cite{Price2008}, superdeterminism~\cite{tHooft2007}, or other mechanisms), but, as yet, no such theory has been proposed that reproduces all the predictions of quantum mechanics.

Finally, it was brought to our attention that the LF polytopes have been independently studied
under the name of ``partially deterministic polytopes''\cite{Woodhead2014}, from an information-theoretic motivation: they are connected to the problem of device-independent randomness certification (see, for example, refs.~\cite{Colbeck2009,PironioNature2010,AcinNature2016} and references therein) in the presence of no-signalling adversaries. 

\section{Methods}

\subsection{Quantum mechanical description of the EWFS}
We consider two superobservers, Alice and Bob, and their respective friends, Charlie and Debbie. Charlie and Debbie are in possession of systems $S_A$ and $S_B$ respectively, with associated Hilbert spaces $\mathcal{H}_{S_A}$ and $\mathcal{H}_{S_B}$, and initially prepared in a (possibly entangled) state $\rho_{S_AS_B}$. For simplicity, we suppose these systems are spin-1/2 particles, and that they perform a measurement of the $z$-spin of their particles. We denote everything in Charlie's lab except $S_A$ as system $F_A$, with Hilbert space $\mathcal{H}_{F_A}$, and $F_B$, $\mathcal{H}_{F_B}$ for Debbie's lab. 

According to Alice, Charlie's measurement of $S_A$ in the basis $\{\ket{-1}_{S_A},\ket{+1}_{S_A}\}$ is described by a unitary $U_{Z_A}$ acting on $\mathcal{H}_{F_A}\otimes\mathcal{H}_{S_A}$. Alice's $x=1$ measurement (corresponding to opening the box and asking Charlie what he saw) can be described by a POVM $\{\proj{c}{c}_{F_A}\otimes I_{S_A}\}_c$, where  $\ket{c}_{F_A}$ ($c \in \{-1,+1\}$) represents the state of Charlie after seeing outcome $c$ and $I_{S_A}$ is the identity operator on $\mathcal{H}_{S_A}$. 
The theorem makes no assumption about the form of the measurements that Alice performs for $x\in\{2,3 \}$, but in our experimental realization, we consider the class of measurements that reverse the evolution $U_{Z_A}$ that entangled $F_A$ with $S_A$ (Fig.\ \ref{fig:concept_comic}b), followed by a measurement on $S_A$ alone (Fig.\ \ref{fig:concept_comic}c). This can be described by a POVM with elements $U_{Z_A}(I_{F_A}\otimes E_{S_A}^{a|x})U_{Z_A}^{-1}$, where $I_{F_A}$ is the identity on $\mathcal{H}_{F_A}$ and $E_{S_A}^{a|x}$ is the positive operator associated with outcome $a$ for measurement $x$ that Alice performs directly on $S_A$.
Bob's POVM elements  
are defined analogously. Thus, the maximum violation of the inequalities can be sought simply in measurements acting on the Hilbert spaces  $\mathcal{H}_{S_A}$ and $\mathcal{H}_{S_B}$; since Charlie and Debbie start in a known product state in the Hilbert space of $\mathcal{H}_{F_A}\otimes\mathcal{H}_{F_B}$, there is no advantage in considering arbitrary measurements on $\mathcal{H}_{F_A}\otimes\mathcal{H}_{S_A}$ and $\mathcal{H}_{F_B}\otimes\mathcal{H}_{S_B}$.

\subsection{LHV correlations as a subset of LF correlations}\label{subset_section}

Recall that a set of correlations has a LHV model if and only if there exists a probability distribution $P(\lambda)$ over a set of variables $\lambda \in \Lambda$ such that
\begin{equation}
\label{separable}
\wp(ab|xy)=\Sigma_{\lambda\in\Lambda} P(a|x\lambda)P(b|y\lambda)P(\lambda),
\end{equation}
for all values of the variables $a,b,x,y$. We now derive the general form for an LF model. From AOE and NSD, we have that:
\begin{eqnarray}
\label{ProbFirst}
\wp(ab|xy) {\stackrel{\text{{AOE}}}{=}} \sum_{c,d}P(abcd|xy) 
         {\stackrel{\text{{NSD}}}{=}} \sum_{c,d}P(ab|cdxy)P(cd).
\end{eqnarray}
From {Locality}, we can decompose the first term on the right-hand side in two ways:
\begin{eqnarray}
\label{Locality-A}
P(ab|cdxy)= P(a|bcdxy)P(b|cdxy) 
        { \stackrel{\text{{L}}}{=} } P(a|bcdxy)P(b|cdy)
\end{eqnarray}
or
\begin{eqnarray}
\label{Locality-B}
P(ab|cdxy)= P(a|cdxy)P(b|acdxy) 
        { \stackrel{\text{{L}}}{=} } P(a|cdx)P(b|acdxy).
\end{eqnarray}
Note, however, that we cannot further reduce these expressions with {Locality} alone---reinforcing the fact that {Locality} is a weaker assumption than local causality (which leads to a LHV model). However, by construction, when $x=1$ we have $a=c$, and when $y=1$, $b=d$. Then, if $x=1$, $P(a|bcdxy)=\delta_{a,c}$, and if $y=1$, $P(b|acdxy)=\delta_{b,d}$. When taking this, along with Eqs.~(\ref{Locality-A}) and (\ref{Locality-B}), into account, we obtain from Eq.~(\ref{ProbFirst})
\begin{equation}
\label{LFgeneral}
\wp(ab|xy)=\begin{cases}
\sum_{c,d}\delta_{a,c}P(b|cdy)P(cd) & \text{ if } x=1 \\ 
\sum_{c,d}\delta_{b,d}P(a|cdx)P(cd) & \text{ if } y=1 \\ 
\sum_{c,d}P_{\mathrm{NS}}(ab|cdxy)P(cd) & \text{ if } x\neq1, y\neq1
\end{cases},
\end{equation}
where $P_{\mathrm{NS}}(ab|cdxy)$ denotes some joint probability distribution that satisfies the condition of {Locality}. For any fixed values of $c$ and $d$, it is easy to see that the set of $P_{\mathrm{NS}}(ab|cdxy)$ is simply the no-signalling polytope with one less measurement setting for both Alice and Bob~\cite{Barrett05} (thus the NS subscript). In general, because of the additional structure given by the first two lines of Eq.~\eqref{LFgeneral}, the set of LF correlations only forms a subset of the no-signalling polytope.

To see that LHV correlations are also LF correlations, we first recall from ref.~\cite{BrunnerRMP} that correlations of the form of Eq.~\eqref{separable} can always be decomposed in terms of the extreme points of the set of such correlations. To this end, it is expedient to write the hidden variable as $\lambda=(\lambda^A_{1},\lambda^B_{1},\lambda^A_{2},\ldots \lambda^B_{N})$, with $\lambda^A_{x}$ and $\lambda^B_{y}$ parameterizing all possible local deterministic strategies, \ie, 
\begin{equation}
\label{determinisim}
	P(a|x\lambda)=\delta_{a,\lambda^A_{x}},\,\,
	P(b|y\lambda)=\delta_{b,\lambda^B_{y}}.
\end{equation}
We may now rewrite Eq.~\eqref{separable} as:
\begin{equation}
\label{decomposition}
	\wp(ab|xy)=\sum_{\lambda} \delta_{a,\lambda^A_{x}}\,\delta_{b,\lambda^B_{y}} P(\lambda).
\end{equation}
This is now readily cast in the form of Eq.~\eqref{LFgeneral} if we set $\lambda^A_{1}=c$ and $\lambda^B_{1}=d$. For example, if $x=1$, we get
\begin{equation}
\begin{split}
	\wp(ab|x=1,y)&=\sum_{c,d,\lambda^A_2,\ldots,\lambda^B_N} \delta_{a,c}\,\delta_{b,\lambda^B_{y}} P(cd\lambda^A_2\ldots\lambda^B_N),\\
		     &=\sum_{c,d,\lambda^B_y} \delta_{a,c}\,\delta_{b,\lambda^B_{y}} P(cd\lambda^B_y),\\
		     &=\sum_{c,d} \delta_{a,c}\,\left[\sum_{\lambda^B_y} \delta_{b,\lambda^B_{y}} \,P(\lambda^B_{y}|cd)\right] P(cd),\\
		     &=\sum_{c,d} \delta_{a,c}\,P(b|cdy) P(cd),
\end{split}
\end{equation}
which is clearly of the form given in the first line of Eq.~\eqref{LFgeneral}. The proof for the $y=1$ case is completely analogous. 

Similarly, for the case where $x\neq1$, $y\neq 1$, we can again make use of $\lambda^A_{1}=c$, $\lambda^B_{1}=d$ and Eq.~\eqref{decomposition} to arrive at:
\begin{equation}\label{LHV->LF}
\begin{split}
	\wp(ab|xy)&=\sum_{c,d,\lambda^A_2,\ldots,\lambda^B_N} \delta_{a,\lambda^A_{x}}\,\delta_{b,\lambda^B_{y}} P(cd\lambda^A_2\ldots\lambda^B_N),\\
		     &=\sum_{c,d,\lambda^A_x\lambda^B_y} \delta_{a,\lambda^A_{x}}\,\delta_{b,\lambda^B_{y}} P(cd\lambda^A_{x}\lambda^B_y),\\
		     &=\sum_{c,d} \left[\sum_{\lambda^A_{x},\lambda^B_y} \delta_{a,\lambda^A_{x}}\,\delta_{b,\lambda^B_{y}} \,P(\lambda^A_{x}\lambda^B_{y}|cd)\right] P(cd),\\
		     &=\sum_{c,d} \,P(ab|cdxy) P(cd).
\end{split}
\end{equation}
From the second last line of Eq.~\eqref{LHV->LF} and the fact that $a$ ($b$) is entirely decided by $\lambda^A_x$ ($\lambda^B_y$),  we see that $P(ab|cdxy)$ in the last expression satisfies the condition of {Locality} (i.e., $\sum_a P(ab|cdxy)$ does not depend on $y$ while $\sum_b P(ab|cdxy)$ does not depend on $x$). Thus, starting from LHV correlations for $x\neq1$, $y\neq1$, we recover the last line of Eq.~\eqref{LFgeneral}.

Hence any correlation that satisfies Eq.~(\ref{separable}) will also satisfy Eq.~(\ref{LFgeneral}). Yet, the opposite is not necessarily true. Therefore, LHV correlations are a subset of LF correlations.

\subsection{Characterization of LF correlations}\label{characterization_section}
Consider a general scenario with $N$ measurement settings per party, with $O$ outcomes each. Note that we can always rewrite Eq.~\eqref{LFgeneral} in the form 
\begin{equation}
\label{LFgeneralDet}
\wp(ab|xy)=\begin{cases}
\sum_{\lambda}\delta_{a,c(\lambda)}\Pext^{(j(\lambda))}(b|y)P(\lambda) & \text{ if } x=1 \\ 
\sum_{\lambda}\delta_{b,d(\lambda)}\Pext^{(j(\lambda))}(a|x)P(\lambda) & \text{ if } y=1 \\ 
\sum_{\lambda}\Pext^{(j(\lambda))}(ab|xy)P(\lambda) & \text{{otherwise}}
\, ,
\end{cases}
\end{equation}
where $\lambda$ is a variable that determines the values of $c(\lambda)$, $d(\lambda)$, and that of a variable $j(\lambda)$ that labels the (finitely many) extreme points of the no-signalling polytope with $N-1$ inputs and $O$ outputs per party, and $\Pext^{(j)}(a|x)=\sum_b \Pext^{(j)}(a,b|x,y)$ and $\Pext^{(j)}(b|y)=\sum_a \Pext^{(j)}(a,b|x,y)$ are the marginal distributions of these extremal boxes. 

It is easy to see from the above that this set of correlations is convex. That is, for any two points $\wp_1(ab|xy)$ and $\wp_2(ab|xy)$, both satisfying the LF conditions, any convex combination $\wp'(ab|xy)=\alpha \wp_1(ab|xy)+(1-\alpha)\wp_2(ab|xy)$, with $0<\alpha<1$, also satisfies those conditions. The set of LF correlations is therefore a polytope.

For the two-measurement-setting case ($N=2$), the $\Pext^{(j)}(a,b|x,y)$ now refer only to the case $x=y=2$, and the extreme points are now simply deterministic functions for $a,b$. Thus, we recover an LHV model for any value of $O$, yielding the same inequalities Brukner derived for $N=O=2$.

Next, we consider the LF polytope for the $N=3,O=2$ scenario. Without loss of generality, we label the outcomes as $a,b \in \{+1,-1\}$. From Eq.~\eqref{LFgeneralDet}, the set of LF correlations $\vecP = \{\wp(a,b|x,y)\}_{a,b=\wp1,x,y=1,2,3}$ is the convex hull of the extreme points $\{\vecP^{(\lambda)}(a,b|x,y)\}_\lambda$ defined by
\begin{equation}
	P^{(\lambda)}(a,b|x,y) = 
	\left\{\begin{array}{r@{\quad \quad}}
		\delta_{a,c(\lambda)}\delta_{b,d(\lambda)}: x=y=1 \\
		\delta_{a,c(\lambda)}\Pext^{(j(\lambda))}(b|y): x=1,y\neq 1 \\
		\Pext^{(j(\lambda))}(a|x)\delta_{b,d(\lambda)}: x\neq 1,y= 1 \\
		\Pext^{(j(\lambda))}(a,b|x,y): x\neq 1,y\neq 1
		 \end{array}\right. .
\end{equation} 

Since there are four combinations of $(c,d)$ corresponding to $2^2$ local deterministic strategies for the first inputs, and 24 extreme points for the aforementioned no-signalling polytope~\cite{Barrett05}, we thus end up with 96 points in this set.

By writing the components of these points in a text file and feeding the latter into the freely available software PANDA---which allows one to transform between the two representations of a polytope using the {\em parallel adjacency decomposition algorithm}~\cite{PANDA}---we obtain the complete set of 932 LF facets for this scenario. Many of these inequalities can be transformed from one to another under a relabeling of parties (Alice $\leftrightarrow$ Bob), inputs ($x=2\leftrightarrow x=3$ and/or $y=2\leftrightarrow y=3$), and/or outputs ($a=+1\leftrightarrow a=-1$ and/or $b=+1\leftrightarrow b=-1$).  With the exception of the settings for $x=1$ and $y=1$, the rest of these labelings are arbitrary. Taking advantage of this arbitrariness, we may group the obtained facets into the following 9 inequivalent classes (written in terms of correlators, where $A_i$ is a random variable representing the measurement result for $x=i$ and taking values $\{-1,+1\}$; similarly for $B_j$):
\begin{enumerate}
\item Genuine LF facet 1 (appearing 256 times among the 932 facets):
	\begin{equation}\label{eq:LF1}
	\begin{split}
	 - \langle A_1 \rangle - \langle A_2 \rangle - \langle B_1 \rangle - \langle B_2 \rangle \\
	     - \langle A_1B_1 \rangle - 2\langle A_1B_2 \rangle  - 2\langle A_2B_1 \rangle + 2\langle A_2B_2 \rangle\\ - \langle A_2B_3 \rangle - \langle A_3B_2 	\rangle - \langle A_3B_3 \rangle -6 \overset{LF}{\le}  0
	\end{split}
	\end{equation}     
\item Genuine LF facet 2 (appearing 256 times):
	\begin{equation}\label{eq:LF2}
	\begin{split}
	- \langle A_1 \rangle - \langle A_2 \rangle - \langle A_3 \rangle - \langle B_1 \rangle \\
- \langle A_1B_1 \rangle - \langle A_2B_1 \rangle  - \langle A_3B_1 \rangle - 2\langle A_1B_2 \rangle\\
 + \langle A_2B_2 \rangle + \langle A_3B_2 \rangle - \langle A_2B_3 \rangle + \langle A_3B_3 \rangle -5 \overset{LF}{\le} 0
	\end{split}
	\end{equation}     
	\item Bell $I_{3322}$~\cite{CollinsGisin} with marginals over input 1 and 2 (appearing 256 times):
	\begin{equation}\label{eq:I3322_1}
	\begin{split}
     - \langle A_1 \rangle + \langle A_2 \rangle + \langle B_1 \rangle - \langle B_2 \rangle \\
    + \langle A_1B_1 \rangle - \langle A_1B_2 \rangle  - \langle A_1B_3 \rangle - \langle A_2B_1 \rangle \\
    + \langle A_2B_2 \rangle - \langle A_2B_3 \rangle - \langle A_3B_1 \rangle - \langle A_3B_2 \rangle -4 \overset{LF}{\le} 0
	\end{split}
    \end{equation}
\item Bell $I_{3322}$ with marginals over input 2 and 3  (appearing 64 times):
	\begin{equation}
	\begin{split}
	 - \langle A_2 \rangle - \langle A_3 \rangle - \langle B_2 \rangle - \langle B_3 \rangle \\
	- \langle A_1B_2 \rangle  + \langle A_1B_3 \rangle - \langle A_2B_1 \rangle - \langle A_2B_2 \rangle\\
	 - \langle A_2B_3 \rangle + \langle A_3B_1 \rangle 	- \langle A_3B_2 \rangle - \langle A_3B_3 \rangle -4 \overset{LF}{\le} 0
	\end{split}
	\end{equation}     
\item ``Brukner inequality'': Bell-CHSH for input 1 and 2 of Alice, and input 1, 3 of Bob (appearing 32 times):	
	\begin{equation} \label{eq:Brukner}
	\begin{split}
\langle{A_1B_1}\rangle-\langle{A_1B_3}\rangle-\langle{A_2B_1}\rangle-\langle{A_2B_3}\rangle-2\overset{LF}{\le} 0
 	\end{split}
	\end{equation}     
\item ``Semi-Brukner'' inequality: Bell-CHSH for input 2, 3 of Alice, and input 1, 2 of Bob (appearing 32 times):
	\begin{equation}\label{eq:half-Brukner}
	\begin{split}
-\langle{A_1B_2}\rangle+\langle{A_1B_3}\rangle-\langle{A_3B_2}\rangle-\langle{A_3B_3}\rangle-2 \overset{LF}{\le} 0
	\end{split}
	\end{equation} 
\item Positivity for input 1 of Alice and input 1 of Bob (appearing 4 times):		    
	\begin{equation}
     1+\langle A_1 \rangle+\langle B_1 \rangle+\langle A_1B_1 \rangle \ge 0
	\end{equation} 
\item Positivity for input 1 of Alice and input 2 of Bob (appearing 16 times):
	\begin{equation}
     1+\langle A_1 \rangle+\langle B_2 \rangle+\langle A_1B_2 \rangle \ge 0
	\end{equation} 
\item Positivity for input 2 of Alice and input 2 of Bob (appearing 16 times):		    	
	\begin{equation}
     1+\langle A_2 \rangle+\langle B_2 \rangle+\langle A_2B_2 \rangle \ge 0
	\end{equation} 
\end{enumerate}

Note that some Bell facets for this scenario are not facets of LF and thus do not appear in the list above, \eg, the Bell-CHSH inequalities that do not include any input 1 for either party:
\begin{equation}
\label{eq:Bell-non-LF}
\langle{A_2B_2}\rangle-\langle{A_2B_3}\rangle-\langle{A_3B_2}\rangle-\langle{A_3B_3}\rangle-2 \overset{LHV}{\le} 0
\end{equation}

\subsection{Inequalities and measurements considered in the experiment}

For each category, the inequalities we considered in our experiment were
``genuine LF'', Eq.~\eqref{eq:LF1}, ``$I_{3322}$'', Eq.~\eqref{eq:I3322_1}, 
``Brukner'', Eq.~\eqref{eq:Brukner}, ``semi-Brukner'', Eq.~\eqref{eq:half-Brukner}, and ``Bell non-LF'', as given by Eq.~\eqref{eq:Bell-non-LF}.

Here we use $A_x\in \{+1,-1\}$ as the random variable for Alice's outcome $a$ when she chooses setting $x$, and similarly $B_y$. That is, the expectation values are calculated from the empirical probabilities $\wp(ab|xy)$.

 We restrict ourselves to projective measurements in the $XY$ plane of the Bloch sphere (with states $\ket{H}$ and $\ket{V}$ on the $z$-axis). In particular, Alice's measurement results are represented by operators of the form $A_x = 2\Pi_x^{a=1} -\ket{H}\bra{H}-\ket{V}\bra{V}$, with $\Pi^{a=1}_x=\ket{\phi_x}\bra{\phi_x}$ being the projector onto the state
\begin{equation}
    \ket{\phi_x} =\frac{1}{\sqrt{2}}\left(\ket{H} + e^{i\phi_x}\ket{V}\right)\,.
\end{equation}
Bob's corresponding operators are chosen to be $B_y = 2\Pi^{b=1}_y-\ket{H}\bra{H}-\ket{V}\bra{V}$, with $\Pi^{b=1}_ y=\ket{\beta_y}\bra{\beta_y}$ being the projector onto
\begin{equation}
    \ket{\beta_y} = \frac{1}{\sqrt{2}}\left(\ket{H} + e^{i(\beta-\phi_{y})}\ket{V}\right)\,.
\end{equation}

 For each value of the tetrad $(\phi_1, \phi_2, \phi_3, \beta)$, and for each category in Table \ref{table}, we find the smallest value of $\mu$ for which one of the inequalities in that category is violated. We then pick a tetrad that makes the gap between these values of $\mu$ conveniently large. The values we choose are $\phi_1=168^\circ$, $\phi_2=0^\circ$, $\phi_3=118^\circ$ and $\beta=175^\circ$. For the inequality in each category that is violated first, we display the values of the left-hand side as a function of $\mu$ in Fig.~\ref{fig:results}.

\subsection{Spontaneous parametric down-conversion source}
The source is made up of an imbalanced pump-beam interferometer (one arm of the interferometer is longer than the other) and two orthogonally oriented (sandwiched) bismuth triborate (BiBO) crystals~\cite{altepeter05}, which are pumped by a 404 nm continuous wave laser diode to produce spontaneous parametric down-conversion. The relative pump power in the interferometer arms determines the $\mu$ parameter of the state and is controlled by the half-wave plate (HWP) after the laser. When all the pump power is in the short arm, the first term of the quantum state, the singlet state, is generated (after a local polarization rotation in the fiber). Conversely, when all the pump power is in the long arm, only the second term, a mixed state, is generated. The beams in both arms are recombined in the non-polarizing beam splitter to pump the sandwiched crystal, generating the desired quantum state.

The polarization in the short arm is rotated to diagonal by a HWP and an additional birefringent element is used to pre-compensate the temporal walk-off in the down-conversion. The polarization in the long arm is also rotated to diagonal by a HWP and a birefringent crystal decoheres the horizontal and vertical polarization components completely, which is necessary to generate the mixed part of the state.

\subsection{Quantum state tomography}

To allow tomography, the motorized mirrors are moved out of the beam paths and the measurements are carried out using the last quarter-wave plate (QWP), half-wave plate (HWP), and polarizing beam splitter (PBS) on each side. As part of the tomographic state reconstruction, the known unitary transformations of the first QWP and the beam displacer (BD) interferometer are accounted for, such that the quantum state straight after the fiber is obtained.  Typically, about 22000 coincidences are collected per tomography.  
The implemented $\mu$ values are estimated by comparing the reconstructed states with the set of target states $\rho_{\mu}$, and finding the $\mu$ values that maximize the fidelity (for details, see Supplementary Information, Sec.\ D.) 

\subsection{Experimental implementation of measurements}

The measurements of the EWFS are realized in the following way. When measurement setting 1 is chosen, the motorized mirror is inserted to reveal the photon path within the interferometer,  i.e.\ after Beam Displacer 1, and this corresponds to Alice asking Charlie his measurement outcome (or Bob asking Debbie on the other side).  This comprises the first of the possible POVMs, illustrated in Fig.\ \ref{fig:concept_comic}a. When measurement settings 2 or 3 are chosen, Alice (Bob) first reverses Charlie's (Debbie's) measurement (Fig.\ \ref{fig:concept_comic}b) by removing the mirror between the two beam displacers and thereby closing the interferometer, and then proceeds to measure the polarization after the interferometer with the QWP after BD2 removed (Fig.\ \ref{fig:concept_comic}c).  This two-step procedure corresponds to Alice (Bob) implementing one of her (his) other two POVMs, depending on which one of two settings of the last HWP is used. Single photons are detected with avalanche photodiodes (APDs) and coincidences are recorded with counting modules. The overall observed rate of counts in the apparatus is approximately 550 coincidences and 21000 singles per second.

To obtain the expectation values required for the inequalities being tested at each $\mu$ value, we performed the nine sets of measurements that arise from combining the three independent measurement settings on Alice's and Bob's sides. The typical number of counts per measurement set is 91000 coincidences. 

\noindent {Acknowledgements:} 
This work was supported by the Australian Research Council (ARC) Centre of Excellence CE170100012, the Ministry of Science and Technology, Taiwan (Grants No.\ 107-2112-M-006-005-MY2 and 107-2627-E-006-001), ARC Future Fellowship FT180100317, and grant number FQXi-RFP-1807 from the Foundational Questions Institute and Fetzer Franklin Fund, a donor advised fund of Silicon Valley Community Foundation. A.U.-A., K.-W.B. and F.G. acknowledge financial support through Australian Government Research Training Program Scholarships, and N.T. acknowledges support by the Griffith
University Postdoctoral Fellowship Scheme. We gratefully acknowledge Antonio Ac\'{\i}n for bringing ref.~\cite{Woodhead2014} to our attention, and thank Sergei Slussarenko for useful discussions. Avatars in Figs.~\ref{fig:concept_fig} and \ref{fig:concept_comic} are adapted from Eucalyp Studio, available under Creative Commons (Attribution 3.0 Unported),  \href{https://creativecommons.org/licenses/by/3.0/}{https://creativecommons.org/licenses/by/3.0/}, at \href{https://www.iconfinder.com/iconsets/avatar-55}{https://www.iconfinder.com/iconsets/avatar-55}.

\newpage

\setcounter{equation}{0}
\renewcommand{\theequation}{S.\arabic{equation}}

\setcounter{figure}{0}
\renewcommand{\thefigure}{S.\arabic{figure}}

\setcounter{table}{0}
\renewcommand{\thetable}{S.\Roman{table}}

\setcounter{section}{0}
\renewcommand{\thesection}{S.\arabic{section}}

\begin{center}
    \Large {\textbf{Supplementary Information}}\ \\
    \normalsize
\end{center}

\section{A. ~Wigner's friend thought experiment}
In the Wigner's friend thought experiment \cite{Wigner} an observer, whom we call the friend, performs a measurement on a quantum system $S$. The friend is in a laboratory that can be coherently controlled by a second experimenter called Wigner. As a superobserver, Wigner has the ability to implement arbitrary quantum operations on the friend's laboratory and everything it contains.

Wigner initially assigns a product quantum state $\ket{\phi_0}_F \otimes \ket{\psi_0}_{S}$  to the overall system composed of the friend, $F$, and the system, $S$. For example, the system may be a spin-1/2 particle, and the friend measures the operator corresponding to spin projection along the $z$ direction, with eigenstates $\ket{\uparrow}_S$ and $\ket{\downarrow}_S$.

From Wigner's perspective, the friend's measurement in the $z$ basis is described as a unitary evolution $U_{Z}$ that correlates the friend (and the display on her measurement apparatus, etc.) to system $S$ in the appropriate way. That is, if the initial state of $S$ is $\ket{\uparrow}_S$, the final state of the joint system is $U_Z(\ket{\phi_0}_F \otimes \ket{\uparrow}_S) = \ket{\mathrm{up}}_F \otimes \ket{\uparrow}_S$, and likewise $U_Z(\ket{\phi_0}_F \otimes \ket{\downarrow}_S) = \ket{\mathrm{down}}_F \otimes \ket{\downarrow}_S$.

An interesting scenario occurs when $S$ is prepared in a superposition state, for example $\frac{1}{\sqrt{2}}(\ket{\uparrow}_S+\ket{\downarrow}_S)$. Then standard textbook quantum mechanics predicts that the friend will observe one or another outcome with equal probability, and the state of the system after measurement (and that of the friend) will be one or another of the corresponding states above. On the other hand, due to the linearity of the unitary map, from Wigner's perspective the final joint state will be $\ket{\Phi^+}_{FS} = \frac{1}{\sqrt{2}}(\ket{\mathrm{up}}_F\ket{\uparrow}_S+\ket{\mathrm{down}}_F\ket{\downarrow}_S)$. This entangled state does not assign well-defined values to the states of $S$ or $F$ separately, and therefore seems to be in direct contradiction with standard textbook quantum mechanics. This contradiction is called the \emph{measurement problem}.

Indeed, if Wigner had the control over $F$ that quantum mechanics in principle allows, then he could measure the POVM $\{\ket{\Phi^+}\bra{\Phi^+}_{FS}, I_{FS} - \ket{\Phi^+}\bra{\Phi^+}_{FS}\}$, and he would always get the outcome corresponding to state $\ket{\Phi^+}_{FS}$, confirming Wigner's state assignment. Had the state of $FS$ before this measurement been an equal mixture of the post-measurement states $\ket{\mathrm{up}}_F \otimes \ket{\uparrow}_S$ and $\ket{\mathrm{down}}_F \otimes \ket{\downarrow}_S$, Wigner would have obtained, with equal probability, either of the above outcomes.

The contradiction arises from the assumptions that (i) quantum theory is universal and can be applied at any scale, even to a macroscopic observer, and that (ii) there is an objective collapse after a measurement~\cite{Formalisms}. Thus no contradiction arises if quantum mechanics does not describe objects as large as the friend, or if the collapse of system $S$ is not an objective physical process affecting the wavefunction described by Wigner.

The latter case poses new questions, however. If wavefunction collapse is not objective, is there nevertheless an objective fact corresponding to the friend's observed outcome? Our Theorem 1 demonstrates a contradiction between the (metaphysical) assumptions of \emph{{No-Superdeterminism}}, \emph{{Locality}} and \emph{{Absoluteness of Observed Events}}, and the (empirical) hypothesis that quantum mechanics is valid, and in principle allows coherent operations (such as the above measurements by Wigner) to be implemented, on the scale of a friend $F$.

\section{B. ~Maximal quantum violations of the Genuine LF inequalities}\label{App:violations}

By implementing a see-saw type algorithm (see, \eg, refs.~\cite{See-saw1,See-saw2,Liang:2007} and references therein), one finds that the Genuine LF inequality 1 \eqref{eq:LF1}, with an LF upper bound of 0, can be violated by quantum correlations up to 1.345 using a partially entangled two-qubit state (with Schmidt coefficients approximately given by 0.776 and 0.631) and rank-1 projective measurements. Moreover, it can be verified by solving a converging hierarchy~\cite{NPA2007,NPA2008,QMP} of semidefinite programs that this quantum violation is (within a numerical precision of $10^{-7}$) the maximum allowed in quantum theory. In terms of noise robustness, this quantum strategy can tolerate up to $18.3\%$ of white noise before it stops beating the LF bound.

For Genuine LF inequality 2 \eqref{eq:LF2} (with an LF upper bound of 0), the best quantum violation that we have found is 0.880, which apparently can only be achieved using a partially entangled two-qutrit state (with Schmidt coefficients approximately given by 0.645, 0.570, and 0.509) and a combination of rank-2 and rank-1 projectors in the optimal measurements. As with the case of Genuine LF inequality 1, this quantum violation is provably optimal (within a numerical precision of $10^{-7}$) using the solution obtained from solving some semidefinite programs. The white-noise tolerance of this inequality is somewhat worse than the other Genuine LF inequality, giving approximately $18.0\%$.

\section{C. ~Further information about Figure~\ref{fig:2dslice}}\label{sec:Slice}

\begin{figure}[h]
    \centering
    \includegraphics[width=0.9\linewidth]{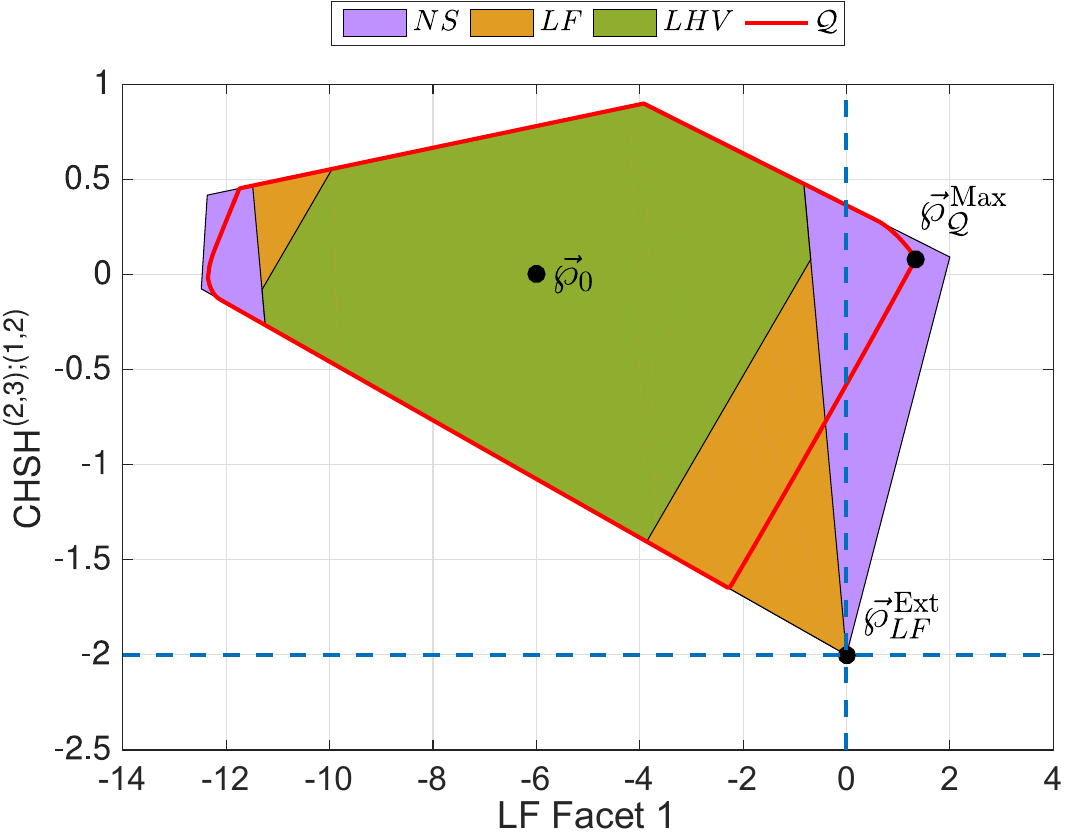}
\caption{\textbf{Detailed version of Fig.\ \ref{fig:2dslice} from the main text.} 
The 2-dimensional slice of the space of correlations is the same as in Fig.\ \ref{fig:2dslice}. This slice is spanned by the three points $\wp_0$, $\wp^{\rm Ext}_{LF}$, and $ \wp^{\rm Max}_{\mathcal{Q}}$, defined in Eqs.\ (\ref{eq:point1}), (\ref{eq:point2}), and (\ref{eq:point3}), respectively. The horizontal axis labels the left-hand side of Eq.~\eqref{eq:LF1} whereas the vertical axis denoted by CHSH$^{(2,3);(1,2)}$ is a short hand for the Bell expression appearing in a Semi-Brukner inequality $-\langle A_2B_1\rangle -\langle A_2B_2\rangle -\langle A_3B_1\rangle + \langle A_3B_2\rangle$. Accordingly, the blue dashed lines demarcate the intersection of the boundary of these facets (each representing a half space) with this 2-dimensional slice. In other words, the LF polytope (even beyond this 2-dimensional slice) has to lie above the horizontal dashed line and to the left of the vertical dashed line.}
    \label{fig:2dslicedetailed}
\end{figure}

Here, we provide further details on the 2-dimensional slice of the space of correlations presented in Fig.~\ref{fig:2dslice}. A variant of this figure containing the same slice, but with further salient features added, is shown in Fig.\ \ref{fig:2dslicedetailed}. Any such 2-dimensional slice is spanned by three affinely-independent correlations in this space (see, \eg, ref.~\cite{Goh18}). In our case, the chosen slice is spanned by the uniform (white-noise) distribution $\vec{\wp}_0$
\begin{equation}\label{eq:point1}
	\wp_0(ab|xy) = \tfrac{1}{4},\quad\forall\, a,b,x,y,
\end{equation}
an extreme point of the LF polytope:
\begin{equation}\label{eq:point2}
\begin{split}
	\wp^{\rm Ext}_{LF}(ab|xy) =
	&\delta_{xy,1}\delta_{a,-1}\delta_{b,1} \\
	+&\tfrac{1}{2}\left[ \delta_{x,1}\delta_{a,-1}(1-\delta_{y,1})+\delta_{y,1}\delta_{b,1}(1-\delta_{x,1}) \right]\\
	+ &\tfrac{1}{4}\left[1+(-1)^{xy-x-y}ab\right](1-\delta_{x,1})(1-\delta_{y,1}),
\end{split}
\end{equation}
 and a symmetrical quantum correlation, written in the Collins-Gisin form (see, \eg, Eq.~(9) of ref.~\cite{CollinsGisin}):
 \begin{equation}\label{eq:point3}
	 \wp^{\rm Max}_{\mathcal{Q}}:
	 \left[\begin{array}{c|ccc}
	  & 0.554 & 0.409 & 0.537\\ \hline
	 0.554 & 0.197 & 0.021 & 0.150\\ 
	 0.409 & 0.021 & 0.311 & 0.040\\ 
	 0.537 & 0.150 & 0.040 & 0.109\\ 
	 \end{array}
	 \right],
 \end{equation}
 \ie, the $i$-th row of the left-most column represent Alice's marginal probability $\wp^{\rm Max}_{\mathcal{Q}}(+1|x=i-1)$, the $j$-column of the top row represent Bob's marginal probability $\wp^{\rm Max}_{\mathcal{Q}}(+1|y=j-1)$, while the remaining entries at the $i$-th and $j$-th column represent the joint probability $\wp^{\rm Max}_{\mathcal{Q}}(+1,+1|x=i-1,y=j-1)$. The quantum correlation $\vec{\wp}^{\rm Max}_{\mathcal{Q}}$ is the one that maximally violates Genuine LF inequality 1 \eqref{eq:LF1}, giving a value of 1.345
 , as explained in Sec.\ B.
 
In our plot, we have chosen the left-hand side of Eq.~\eqref{eq:LF1} to label our horizontal axis , while the vertical axis is labelled by the left-hand side of the Semi-Brukner inequality $-\langle A_2B_1\rangle -\langle A_2B_2\rangle -\langle A_3B_1\rangle + \langle A_3B_2\rangle \ge -2$. Different choices would lead to affine transformations of the plot. Also shown in the figure are a dashed vertical line and a dashed horizontal line intersecting at $\vec{\wp}^{\rm Ext}_{LF}$. These dashed lines mark a projection of the boundary of the LF polytope---as given by inequality~\eqref{eq:LF1} and a relabeling of inequality~\eqref{eq:half-Brukner} to give a lower bound of $-2$ as allowed by LF correlations---on the plane that we have chosen. Note also that the set of LHV correlations (coloured green in the figure) could also touch this boundary of $-2$, but this does not take place on the 2-dimensional plane that we have chosen.

\section{D. ~Experimental quantum states}

 We obtain the experimental quantum states through tomographic state reconstruction based on maximum-likelihood estimation. For each experimental state $\rho_{\mathrm{exp}}$, the highest Uhlmann--Jozsa fidelity{~\cite{Liang:2019aa}} $\left[\mathrm{Tr}\left(\sqrt{\sqrt{\rho_{\mathrm{exp}}}\rho_{\mathrm{\mu}}\sqrt{\rho_{\mathrm{exp}}}} \right)\right]^2$ with the family of states $\rho_{\mu}$ is provided in Table \ref{tab:state quality}, along with the corresponding best $\mu$ value. Uncertainties represent $\pm 1$ standard deviations, estimated based on Monte Carlo simulations using 100 samples of Poisson-distributed photon counts.

\begin{table}[htb]
    \centering
    \begin{tabular}{|c|c|}
        \hline
        $\mu$-parameter   & Fidelity  \\ \hline
        0.992 $\pm$ 0.002 & 0.9789 $\pm$ 0.0007\\ \hline
        0.921 $\pm$ 0.002 & 0.9883 $\pm$ 0.0007\\ \hline
        0.866 $\pm$ 0.002 & 0.9887 $\pm$ 0.0007\\ \hline
        0.809 $\pm$ 0.002 & 0.9868 $\pm$ 0.0007\\\hline
        0.798 $\pm$ 0.002 & 0.9873 $\pm$ 0.0007\\ \hline
        0.744 $\pm$ 0.002 & 0.9824 $\pm$ 0.0007\\ \hline
    \end{tabular}
    \caption{\bf{Characterization of the six experimental states with respect to the family of target states.}}
    \label{tab:state quality}
\end{table}

\end{document}